# Challenges and Opportunities of Future Rural Wireless Communications

Yaguang Zhang, David J. Love, James V. Krogmeier, Christopher R. Anderson, Robert W. Heath, and Dennis R. Buckmaster

*Abstract*—Broadband access is key to ensuring robust economic development and improving quality of life. Unfortunately, the communication infrastructure deployed in rural areas throughout the world lags behind its urban counterparts due to low population density and economics. This article examines the motivations and challenges of providing broadband access over vast rural regions, with an emphasis on the wireless aspect in view of its irreplaceable role in closing the digital gap. Applications and opportunities for future rural wireless communications are discussed for a variety of areas, including residential welfare, digital agriculture, and transportation. This article also comprehensively investigates current and emerging wireless technologies that could facilitate rural deployment. Although there is no simple solution, there is an urgent need for researchers to work on coverage, cost, and reliability of rural wireless access.

## I. INTRODUCTION

For the vast majority of broadband users living in urban environments, it can be difficult to understand the imbalance in communication resources between urban and rural regions throughout the world. Network operators prioritize urban tower density over ubiquitous geographic coverage. Considering the network of a U.S. cellular carrier as an example (Fig. 1), large cities have high cell tower counts per 1000 km$^2$, typically over 30. Users in suburban Chicago enjoy 80 to 165 towers per 1000 km$^2$. In sharp contrast, 70.5 percent of Indiana's land has less than ten towers per 1000 km$^2$, while 44.5 percent has less than five.

These disparities are unnoticed by most urban and suburban users. Fig. 1(a) shows that cell towers cluster not only in cities and towns but also along highways. Therefore, even when traveling, most users lack an accurate understanding of the broadband inequality. The National Association of Counties tested the Internet speeds of 3069 U.S. counties and found that over 65 percent were experiencing Internet speeds below the Federal Communications Commission (FCC) broadband definition (25 Mbps download, 3 Mbps upload) [2].

The 1G and 2G cellular eras had the simple objective of providing voice connectivity. Consequently, infrastructure construction based on population density (with large macrocells in rural areas) was an efficient, cost-effective approach. In the U.S., rural regions account for 97 percent of the land area but only 19.3 percent of the population [3]. Achieving broadband connectivity over such a large geographic area requires a high initial investment, as more towers are needed for broadband vs. voice service. For instance, the average cost for constructing one conventional cellular site is estimated to be US$200,000–250,000, which is difficult to recover from a low density of potential rural users [4]. This fundamental revenue problem is arguably the primary culprit for the digital divide.

In 5G and future standard deployments, there will be an increasing demand for the connection of physical objects [5]. Cisco predicts that by 2023, the number of devices connected to IP networks will be over three times the global population [6]. With the shift from connecting people to connecting things, new applications will require rural broadband to sustain the economy. According to the U.S. Department of Agriculture (USDA), digital agriculture could drive an annual additional gross benefit of US$47–65 billion, corresponding to nearly 18 percent of annual agricultural production in 2017, and rural broadband connectivity could contribute over one-third of this [7]. The digital divide prevents such visions from being realized. Furthermore, broadband access has become a necessity instead of a luxury, especially during and after the COVID-19 pandemic. The digital divide is causing inequality in multiple dimensions, which could economically and socially cripple rural communities without intervention [8].

Promoting rural broadband and closing the digital gap have been top priorities of the USDA [7], the FCC [9], and the National Institute of Standards and Technology (NIST) [10]. However, the telecom industry still focuses on better serving areas with higher population densities. The digital gap has also been widened by 5G technologies such as millimeter-waves (mmWaves) requiring dense infrastructure deployment. The growing vision of connecting everything will only make ubiquitous coverage increasingly important. It is therefore crucial for the communications research community to achieve higher data rates in rural areas and develop innovative technical solutions to drive down the cost of rural networks.

This article examines the motivations and challenges of providing broadband access over vast rural regions, with an emphasis on wireless communication. An application-centric attitude was applied to reveal various benefits of bridging the digital divide. Key research goals were clarified based on the application requirements and unique features of rural environments. Moreover, this article introduces a comprehensive list of promising rural wireless technologies. In the foreseeable future, rural wireless research will involve developing, improving, and choosing technologies to balance primary trade-offs for each application use case.

## II. MOTIVATIONS AND APPLICATIONS

This section showcases selected wireless applications to demonstrate the benefits of improving rural connectivity.

### A. Overview

Wireless technologies are expected to support multiple future applications in key rural economic sectors (Fig. 2). On the *access network*, the objects of interest in many *outdoor* situations are mobile and/or scattered over a large area. It is normally easier to connect these objects wirelessly.

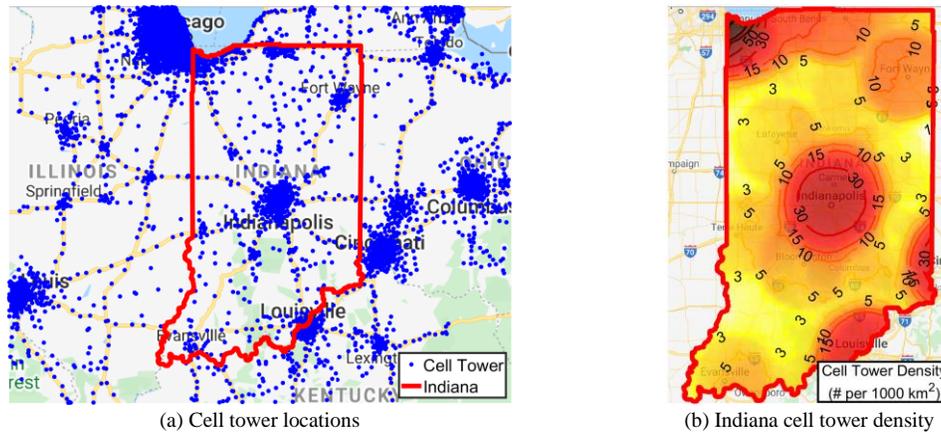

(a) Cell tower locations　　　　　　　　　　　　(b) Indiana cell tower density

Fig. 1. Tower locations for a national cellular carrier according to a randomized real network laydown from the National Telecommunications and Information Administration [1]. Map data ©2021 Google. (a) Cell towers are clustered around densely populated areas. (b) The geographic cell tower density was evaluated within a 50-km radius. Decreasing this radius would yield a map showing a larger digital gap.

For *indoor* applications, the convenience provided by wireless connections is often preferred, especially for personal devices and wearable sensors. At the *backhaul*, optical fiber is the technology of choice for most Internet service providers (ISPs) and wireless communications carriers. The deployment remains relatively expensive, thus wireless alternatives (e.g., *fixed wireless*, *cellular*, and *satellite*) may be considered.

A diverse set of wireless technologies is currently available for different rural communication scenarios. An excellent comparison can be found in [4]. In general, *Bluetooth* suits short-range point-to-point personal-device communications. *Wi-Fi* dominates short/medium-range private networks covering local facilities. *Fixed wireless* can extend backhaul to remote locations. *Bluetooth Low Energy* and *ZigBee* are designed for short-range low-energy applications, while *LoraWAN* and certain *specialized communication networks* for the *Internet of things (IoT*; e.g., SigFox) are for low-throughput, long-range low-energy applications. *Cellular* and *satellite* networks, due to their accessibility, are the de facto options for Internet in remote regions, but with relatively high service costs. *Private wireless networks*, including long-rage/mesh Wi-Fi and private 5G, may be more competitive in the long run in some cases.

Given the wide range of scenarios, there is no one-size-fits-all solution. For the foreseeable future, case-by-case analyses will be needed to choose and orchestrate a mixture of current and emerging technologies that will best balance the desired application trade-offs. Rural wireless will play a vital role, especially for last-mile networks to reach end-user devices [4], and potentially for setting up backhaul at a lower cost in regions with difficult terrains. Furthermore, an abundance of applications is the main factor that will ensure stable rural broadband development. Given the close relationship between wireless technologies and applications, advances in rural wireless research could accelerate application innovation and incubate profitable business models for rural network deployment.

### B. Residential Welfare

Inadequate communication infrastructure prevents rural residents from having the benefits of digital life, including online shopping, mobile banking, social media, and online entertainment. With improved rural broadband, better access to *online services* will combat rural flight by increasing quality of life and expanding economic opportunities [10].

Facilitated by email, instant messaging, videotelephony, and virtual private networks, *telecommuting* has become a feasible, flexible employment option. This can lead to cost savings for both employers and employees by reducing daily commute and office space requirements. Companies could consider candidates for remote jobs with fewer geographic limitations. Future technological advances in augmented reality, virtual reality (VR), and telepresence will further improve the work-from-home experience. However, many of these technologies are only accessible with sufficiently high-speed communication, which is severely lacking in rural areas.

Similarly, *tele-education* provides an alternative to physical presence by delivering content through broadband Internet. Radio and television broadcasting have been used in distance education for several decades. Nevertheless, creating, producing, and distributing educational content is time-consuming, and broadcast communication is only one-way. Real-time two-way communication via the Internet is essential for student engagement and participation. The lack of broadband access shuts many rural residents out of high-quality online education resources that urban residents take for granted.

Many rural areas also lack access to healthcare services. Rural broadband can enable *telehealth* as a solution. For non-emergency services, remote diagnostics allow rural residents to reach healthcare providers with ease. Wearable/implanted medical devices and sensors could monitor chronic health conditions and provide advance warnings of potential problems.

### C. Digital Agriculture

Several key digitalization trends are shaping future methodologies and practices in agriculture. On one hand, *precision agriculture* grants farmers finer control over operations to optimize their production (e.g., the ability to adjust seed placement, fertilizer, and crop protection to centimeter-level accuracy). On the other hand, monitoring technologies are becoming available for both operations and environments; this fosters *smart agriculture*, which exploits data-driven decision-making for improved outcomes.

This new generation of agriculture is among the biggest motivators for accessible broadband and improved wireless technologies in rural areas [7]. As a manufacturing business, agriculture has logistics at its core. Success depends heavily on getting the right thing to the right place at the right time for all kinds of operations, including planting, watering, fertilizing, harvesting, and transporting. Accurate and timely information is required for good decision-making and efficient actions. Wireless communication will become one of *digital agriculture*'s fundamental building blocks throughout production and market coordination stages.

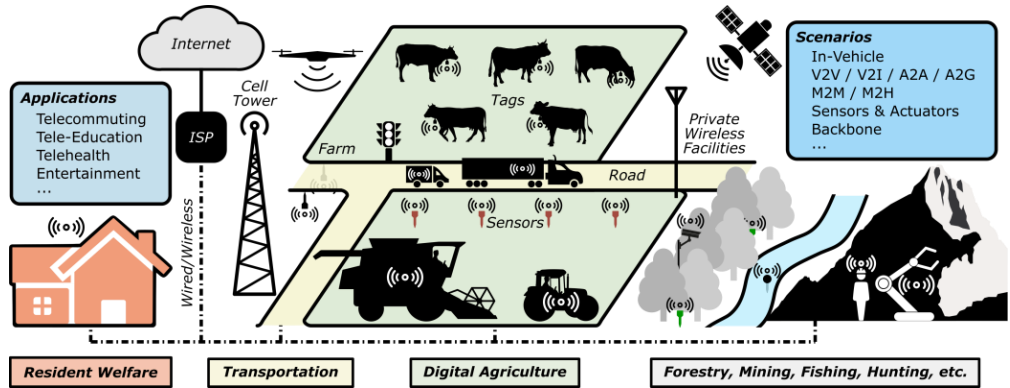

Fig. 2. Illustration of applications and opportunities for future rural wireless communications.

Agricultural logistics depends on an ever-increasing number of *data collection* devices in areas with serious geographic and operational limitations. Fig. 2 shows three such examples: soil condition sensors, livestock identification tags, and unmanned aerial vehicles for field imagery. Other examples include on-site weather stations, vehicle sensors, and mobile assistant apps for human operators. For batch deployment, wired connections and periodic node retrieval quickly become impractical as the number of devices increases. This makes wireless access critical for tasks such as remote control, software deployment, data off-loading, and system upgrades. In many ways, agriculture is at the forefront of widespread IoT applications.

*Communication* equipment dedicated to providing links to IoT and end-user devices is crucial for automating data collection, organization, and distribution, especially for large farms. This category incorporates components of local/wide area networks (LANs/WANs) covering the geographic area of interest. When Internet is available, backhaul is normally constructed and maintained by ISPs and cellular service providers (Fig. 2). Connections directly experienced by users are delivered via satellite, cellular, private wireless facilities, or wired options. Future agricultural vehicles and specialized drones may provide temporary local coverage via data relay.

Ultimately, the value of digital agriculture will be realized through *data-driven decision-making*. The information flow required to close the loop of data collection, decision-making, and action-taking involves wireless technologies at different scales. For instance, the agricultural vehicle fleet in Fig. 2 may be equipped with advanced driver assistance systems or fully autonomous navigation. Sensor-collected vehicle status and environmental information could be shared as needed to avoid collisions and improve cooperation. One way to achieve this is to extend the vehicle bus, the internal network connecting components within the vehicle, with wireless modules. In an edge-computing case, in-vehicle computers gather and analyze data, sharing results with nearby vehicles via vehicle-to-vehicle (V2V) links when necessary. If the decision-making process involved cloud servers, vehicle-to-infrastructure (V2I) communication would be needed for data and command transmission. Alternatively, specialized drone swarms may directly carry out agricultural operations (e.g., planting, spraying, field scouting, soil/plant condition monitoring, and herding) with air-to-air (A2A) and air-to-ground (A2G) communication links.

### D. Transportation

Transportation is undergoing a phase transition from human-driven transportation to full automation. This trend is enabled by advances in sensing, computation, and wireless communication. The same technologies, including machine learning and computer vision, are leading to advancements in agricultural vehicle automation and smart farming. In the long term, this will lead to the integration of smart systems in rural and urban areas. Solving these challenges requires reliable wireless links.

*Safety* is a significant consideration on rural roads. Many intersections are uncontrolled, lacking streetlights or stop signs. Railroad crossings may not have warnings or gates. The composition of vehicles is heterogeneous, including slow-moving farm vehicles and trucks. According to the U.S. Department of Transportation, 49 percent of U.S. car crash fatalities in 2015 occurred in rural regions, despite the low population [11]. Wireless connectivity could reduce these fatalities by 80 percent [11].

In addition to heterogeneous vehicles, rural settings contain diverse roadway types, making *vehicle automation* challenging. There are smaller roads, unpaved roads, and private roads. Livestock, other animals, workers, and manned/unmanned vehicles/equipment may be present. Communication and sensing create opportunities for efficient interactions between vehicles, animals, people, and robots. A vehicle may communicate with other connected entities to coordinate activities and avoid collisions. Alternatively, a vehicle may sense the environment to identify non-connected entities and broadcast this information to other vehicles, increasing situational awareness. Co-locating sensing, computation, and communication at the base station infrastructure multiply the benefits [12]. Data collected from different vehicles could be used to build a dynamic map of the environment including non-connected entities, reducing collisions and improving roadway use efficiency. This is particularly valuable in establishing supply lines between farms/factories and cities.

A great migration from urban to rural areas has been observed across the world since 2020. During the pandemic, people realized that many jobs could be done remotely, removing the need for job proximity. This trend is likely to continue despite in-person requirements as automated vehicles reduce the stress of commuting and make it possible to work, relax, or sleep between home and office. This migration will bring more vehicles and people to rural areas, providing more customers to justify the financial costs of infrastructure expansion.

### E. Other Economic Sectors

Wireless technologies will benefit many other key rural economic sectors—including forestry, mining, fishing, hunting, and manufacturing—in terms of *environmental monitoring* and *emergency detection* (Fig. 2). For heavy mining and manufacturing equipment, the deployment of machine-to-human (M2H) and machine-to-machine (M2M) communication links could be extremely useful for machine/operator *status monitoring*, *remote diagnostics and control*, and *process automation*. Efforts to provide networks catered to specific industrial needs (e.g., hospitals and factories) are observed globally, e.g., local 5G in Japan and Industrial 4.0 in Germany. These wireless networks are deployed to support large-volume communications, enhanced reliability, and/or extremely low latency [13], for use cases such as data collection, surgeries, or operation of industrial machineries.

Reaping the benefits of rural digitalization also involves connecting existing widespread offline infrastructures across economic sectors. For example, in food manufacturing, *material and product traceability* will enable effective product recalls at a minimal cost. This requires automated information sharing along the supply chain through connections between assets owned by different entities, including farmers, ranchers, producers, and manufacturers.

### III. RURAL WIRELESS APPLICATION REQUIREMENTS

Requirements for future rural networks will be at least as stringent as those in urban settings. The greatest technical challenge is to satisfy the diverse requirements of different applications and scenarios with reasonable trade-offs. Fig. 3 illustrates the typical ranges of *throughput* and *latency*, the two most important quality of service (QoS) indicators, for representative applications. Required links vary from low-speed intermittent connections to high-speed stable connections. Collision avoidance via connected vehicles requires V2V links with extremely low latency (one millisecond or less) to limit the response time. In contrast, data from seasonal sensors in agriculture (e.g., soil condition sensors) can endure low data rates and delays of hours to days. To ensure a good user experience, some applications (e.g., video transmission and VR streaming) need data rates at tens of Mbps or higher. The FCC is considering increasing the minimum download speed for broadband from 25 Mbps to 100 Mbps [9].

Other key wireless QoS requirements include *coverage*, *power usage*, *reliability*, and *availability*. Although wide *coverage* is desired in many rural wireless applications, wearable and implanted telehealth sensors may require specialized short-range body area networks to prioritize transmission *reliability* and/or *power consumption*. Some battery-powered devices, such as implantable medical sensors, must operate for long periods between charges. Given that installation is expensive and time consuming, these sensors are expected to be extremely compact and long lasting, leading to *power supply* and wireless module design challenges. Another example is soil sensors, alongside many other IoT devices. Mistake-intolerant applications (e.g., telemedicine and self-driving vehicles) may set high standards for *reliability* and *availability* to control safety risks. Network *latency*, *security*, and *availability* issues will be more prominent in rural areas due to long end-to-end distances.

In practice, non-QoS considerations often dominate the decision-making process for rural network deployment. *Cost* will remain as the primary obstacle to ubiquitous broadband access. *Power supply* via traditional electric grids is unstable or even unavailable in some rural areas. The network must be resilient to extended outages for real-time applications. Also, challenging *deployment environments* with harsh weather or terrain conditions could further increase the difficulty of fulfilling desired QoS requirements. The installation and maintenance of rural communication systems are significantly encumbered by these factors.

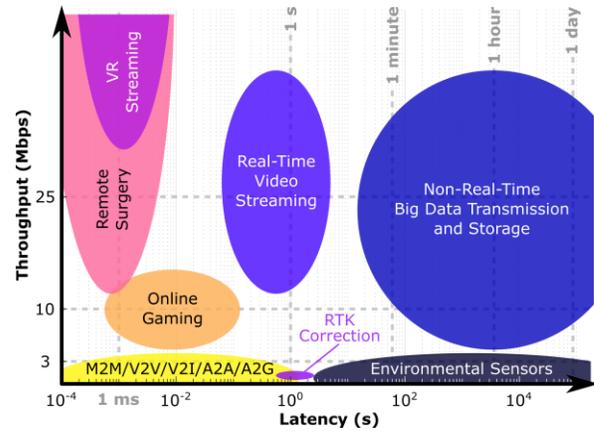

Fig. 3. Typical throughput and latency ranges for selected rural applications. Requirements vary significantly on a case-by-case basis.

Many of the aforementioned factors are intrinsically contradictory. For instance, with other conditions held constant, wider coverage often requires more power consumption, and better QoS performance typically comes at a higher cost. Depending on the scenario, some metrics may warrant prioritization. Besides, the following unique features of rural wireless should be noted:

1) *Large Areas With Limited Communication Resources*
   Backhaul and access network facilities are, and probably will remain, limited in most rural regions.
2) *Highly Dynamic*
   The demand for rural wireless applications is expected to change dramatically both in time and location. For example, in many agricultural activities, the crew and equipment require connection only when and where they are working.
3) *High Demand for Uplink*
   The popularity of data-driven decision-making will remarkably boost data collection and the demand for uplink.
4) *Limited Power Supply and Long Service Time*
   Another infrastructure barrier to rural wireless is unreliable or unavailable power supply, especially for long-service-time wireless devices.
5) *Evolving Vehicular Platforms*
   Vehicles (e.g., for agriculture) are evolving rapidly, with increased computation power and communication ability.
6) *Harsh Deployment Environments*
   Special attention should be paid to deployment environments (e.g., temperature, humidity, and electromagnetic interference).

### IV. FUTURE RESEARCH DIRECTIONS AND CHALLENGES

Cheaper, more efficient methods are still needed to bring backhaul access to rural areas. For future applications with extremely strict QoS requirements, most currently available wireless technologies fall short in terms of performance. Here, we compose a list of research areas that will greatly benefit rural broadband in three loosely defined categories: *extending coverage for existing systems*, *enabling novel low-cost networks*, and *facilitating network construction and maintenance*. The technologies introduced here outline research challenges for future rural wireless communications.

### A. Coverage Extension for Existing Systems

Because of the insufficient rural coverage, there is an urgent need for research on cost-effective wireless techniques to enable backhaul deployment in areas lacking infrastructure. There are multiple ways to approach this challenge, including *free-space optics*, *point-to-point sub-6 GHz radio relaying*, and *point-to-multipoint sub-6 GHz access technologies*. With the use of higher-frequency bands in next-generation wireless networks, extremely high-throughput *fixed wireless* links via *mmWaves* or *terahertz communication* could provide competitive speeds relative to wired options.

In this heterogeneous-band context, regulators could make significant contributions via *spectrum management*, particularly in rural regions because of their low spectrum utilization. *Spectrum sharing* regulated by local policies would potentially alleviate the rural spectrum bottleneck. Research facilitating the administrative decisions will play a key role in balancing the sharp conflict between cost and performance. Low bands are typically better suited for extending coverage in rural areas. Providing large contiguous sub-6 GHz channels could substantially improve both cost effectiveness with cheap radios and network performance with high throughputs. Along this path, cellular carriers are leveraging their low-band assets in rural deployment (e.g., T-Mobile at 600 MHz and China Mobile at 700 MHz).

The long-range communication required in the rural setting will necessitate *multi-hop* approaches. Though multi-hop has had minimal commercial impact until now, it is likely a key technology in any rural communications network. Also, 5G and beyond will depend heavily on mmWave frequencies and multi-hop links to improve their coverage range. Wireless links are also prone to weather-based challenges, typically requiring redundant relaying nodes. Fallback plans (e.g., *satellite access*) may be needed to ensure satisfactory availability. Latency is another distinctive challenge faced by these systems in rural regions, especially for *multi-hop networks*.

*Multiple antenna technologies*, which can be deployed on centralized or distributed platforms (e.g., UAVs, farm equipment, etc.), will extend wireless. *Massive multiple-input multiple-output* (MIMO) and *beamforming* are essential for a successful utilization of mmWaves. C*ooperative MIMO*, another potentially cost-effective technology to enhance rural wireless coverage and capacity, could harvest the theoretical benefits of MIMO at the network level by grouping multiple devices to cooperatively function as virtual antenna arrays. However, there are still numerous challenges involved in managing the increased network complexity and overhead communications required for inter-device cooperation.

### B. Novel Low-Cost Network Enablers

New wireless networks, possibly enabled by disruptive technologies, could also extend rural broadband coverage at a reduced cost. Researchers have been striving for *autonomous networks* with abilities to configure, monitor, maintain, and optimize operations independently. In digital agriculture, intermittent connection established by *store-and-forward networks* is often sufficient. Many operations (e.g., planting, tilling, spraying, and harvesting) require transportation of crews and vehicles between locations with Internet access and fields. This builds a channel for storing, carrying, and forwarding data opportunistically, facilitating retrieval activities for time-insensitive data.

Research challenges in this highly dynamic network environment include security and efficient routing. Nodes can show up and leave. Therefore, optimization in peer discovery, network topology prediction, and path planning may easily exceed limits on node resources, including energy, storage, and computation. Furthermore, communication links are rarely guaranteed. Multiple wireless technologies can be used together to increase connection availability. A*rtificial intelligence* (AI) and *machine learning* may be exploited to take advantage of the nodes' movement, data source, and channel patterns.

*Non-terrestrial networks* (NTNs)—composed of stations at different altitudes carried by *drones*, *high-altitude platforms* (e.g., helikites), and *satellites*—are expected to substantially improve rural broadband. By extending traditional terrestrial networks with aerial and spatial wireless technologies, NTNs could provide cost-effective coverage in unserved areas. Particularly, low-Earth-orbit satellite communications are being implemented and used to deliver rural broadband and cellular connectivity as we speak. The excellent coverage of one satellite, even with limited available bandwidth, perfectly fits the low-density rural scenarios. However, heavily depending on this approach would severely hamper with ground-based astronomy and increase space debris around the earth.

*Drones* are among the most disruptive technologies on our list due to their multi-use nature. Aside from the wireless applications discussed in Section II, extensive rural deployment of drones is expected in many other specialized operations, including aerial imagery/mapping, package delivery, forestry planting, construction progress monitoring, and disaster/public safety response. Drones have the potential to revolutionize rural wireless networks with increased flexibility and 3-dimentional (3D) mobility [14]. *Drone base stations* enable multi-layer 3D network architecture and low-cost on-demand infrastructure deployment, which could mitigate unnecessary operating costs often endured by traditional always-on, fixed infrastructures. *Data relay drones* may play a role in near-future access networks. For example, sending relay drones to follow rural users could enlarge the effective system-level cellular coverage area of Indiana by over 40 percent [14]. This approach is cheaper than satellites and lowers latency via reduced relay altitude at the cost of a smaller coverage area per unit.

Drone technology is still rapidly evolving. Although in outdoor environments drones are more likely to experience line-of-sight channels than users at the pedestrian level, blockages can still happen. Accurately modeling A2A/A2G channels is important for reliability [15], especially in future high-speed networks with mmWave and/or terahertz communications. Drone deployment also faces resource management challenges. Operation time is largely limited by batteries. This can be mitigated via battery capacity improvements, alternative/hybrid energy sources, and alternative/hybrid flight mechanisms (e.g., propellers, wings, and inflatable balloons). Furthermore, wireless communication resource allocation requires optimization to avoid interference. A reliable, low-latency, and possibly low-throughput control link could be critical in managing multiple carrier bands for different QoS levels.

*Topology and trajectory optimization* poses another challenge. The dynamic topology problem of NTNs with drones is different from that of traditional networks, for which the movement of participating nodes is often assumed to be unknown. In contrast, drone locations for NTNs are controllable, and desired network topologies may be proactively constructed via 3D trajectory planning. For large-scale deployment, an increasing number of drones could easily push the complexity of the optimization problem to an impractically unsolvable state.

The success of drones in rural wireless also depends heavily on accurate system-level *performance analyses* via field tests or simulations. These analyses facilitate pre-deployment feasibility studies/post-deployment tests to ensure/verify expected network behaviors. If feedback from these analyses is available in real time, the network could self-adjust its structure via AI to improve performance.

The tremendous number of IoT devices to be connected in rural regions demands a high computational capacity for network management. Agricultural vehicles could function as mobile data hubs/centers to alleviate this burden via *fog/edge computing*. With telematics gaining more attention, these vehicles are changing rapidly to embrace digital technologies. For instance, modern combine harvesters are equipped with various sensors, high-precision GPS, yield monitors, and interactive displays. The *convergence of communication and computing* may also yield innovative, cost-effective solutions for rural wireless applications.

### C. Network Construction and Maintenance Facilitators

Developing technologies to ease network construction and maintenance can reduce the cost of rural broadband expansion. If the same wireless technology is used for both access and backhaul networks, it will be easier to add new infrastructure nodes into existing networks. This technology is called *integrated access and backhaul* and has been included in 5G standards to decrease the cost of dense small cell deployment. It also has huge potential for improving coverage by using lower bands for transport to reduce the number of hops. However, given that resources will be shared between access and backhaul networks, this approach is more challenging than traditional networks in scheduling, load balancing, and interference management.

Research on *network abstraction* improves network management flexibility. For example, *software-defined networking* aims to disassociate routing from data forwarding to create one centralized intelligent control panel, while *network function virtualization* aims to partition the communication service into key functions and encapsulate the hardware devices required to achieve these functions in software. These methods could assist the development and deployment of advanced network techniques (e.g., network slicing). The centralized approach faces security and efficiency challenges, but the resultant abstraction promises programmable and dynamic adjustments, which are essential for automating network operations and incorporating intelligence into resource management. Moreover, knowledge sharing and unsupervised learning among nodes will promote real-time network optimization.

It is unlikely that one existing or near-future technology could satisfy all application needs. The coexistence of wireless technologies will bring about challenges in interference and privacy management. We need an optimized way of utilizing *diverse radio access technologies* over licensed and unlicensed bands to improve rural networks' coverage and performance. Furthermore, research on *low-power nodes*, *energy harvesting*, and *simultaneous wireless information and power transmission* may relieve the energy provision challenge faced by many rural wireless systems. Some research directions not directly related to communications are also of importance. For example, improvements in *battery technology* and *alternative power sources* could benefit communication module design for field sensors, animal tags, drones, and remote base stations.

## V. Discussion

The digital divide is more severe than many realize. Rural wireless will play a crucial role in closing this gap. Advances in wireless technologies will reduce the cost and improve the performance of rural broadband. Future wireless applications may turn rural network construction into a profitable practice. This article comprehensively introduces the challenges, opportunities, future applications, motivations, requirements, and promising research directions of rural wireless. Closing the digital gap will involve policy-making, business development, and research innovation. Efforts and collaborations in government, industry, and academia are required to jointly utilize a mixture of technologies for efficient, cost-effective rural network deployment.

### Acknowledgments

This work was supported by the Foundation for Food and Agriculture Research under Award 534662 and by the National Science Foundation under Grants EEC1941529, CNS1642982, CNS1731658, and CCF1816013.

### References

[1] "Working group 5 final report 1755-1850 MHz airborne operations," National Telecommunications and Information Administration, Commerce Spectrum Management Advisory Committee, Sep. 16, 2013. Accessed: Apr. 4, 2021. [Online]. Available: https://www.ntia.doc.gov/files/ntia/publications/wg5_1755-1850_final_reportl-09-16-2013.pdf

[2] A. Scott. "Understanding the true state of connectivity in America." National Association of Counties, Mar. 2020. https://www.naco.org/resources/featured/understanding-true-state-connectivity-america (accessed Jan. 26, 2021).

[3] "New census data show differences between urban and rural populations." U.S. Census Bureau, Dec. 8, 2016. Accessed: Jan. 26, 2021. [Online]. Available: https://www.census.gov/newsroom/press-releases/2016/cb16-210.html

[4] "The last-mile Internet connectivity solutions guide: Sustainable connectivity options for unconnected sites," International Telecommunication Union, ITU Publications, Geneva, Switzerland, 2020.

[5] Z. Zhang *et al.*, "6G wireless networks: Vision, requirements, architecture, and key technologies," *IEEE Veh. Technol. Mag.*, vol. 14, no. 3, pp. 28–41, 2019, doi: 10.1109/MVT.2019.2921208.

[6] "Cisco annual Internet report (2018–2023)," Cisco Public White Paper, 2020. Accessed: Jan. 26, 2021. [Online]. Available: https://www.cisco.com/c/en/us/solutions/collateral/executive-perspectives/annual-internet-report/white-paper-c11-741490.pdf

[7] "A case for rural broadband: Insights on rural broadband infrastructure and next generation precision agriculture technologies," American Broadband Initiative, U.S. Department of Agriculture, Apr. 2019. Accessed: Jan. 26, 2021. [Online]. Available: https://www.usda.gov/sites/default/files/documents/case-for-rural-broadband.pdf

[8] T. S. El-Bawab, "Toward access equality: Bridging the digital divide," *IEEE Commun. Mag.*, vol. 58, no. 12, pp. 6–7, 2020, doi: 10.1109/MCOM.2020.9311933.

[9] "2020 broadband deployment report," Broadband Progress Reports, Federal Communications Commission, Apr. 24, 2020. Accessed: Jan. 26, 2021. [Online]. Available: https://docs.fcc.gov/public/attachments/FCC-20-50A1.pdf

[10] D. Buckmaster *et al.*, "Rural America, rural economies and rural connectivity," in *Agriculture/Rural Supercluster Blueprint, Global City Teams Challenge*, Washington, DC, U.S., Jul. 10, 2019. Accessed: Jan. 26, 2021. [Online]. Available: https://pages.nist.gov/GCTC/uploads/blueprints/2019-Ag-Rura-Blueprint.pdf

[11] "Intelligent transportation systems (ITS) benefits for rural communities," U.S. Department of Transportation, ITS Professional Capacity Building (PCB) Program, 2018. Accessed: Jan. 26, 2021. [Online]. Available: https://www.its.dot.gov/factsheets/pdf/Rural.pdf

[12] A. Ali, N. Gonzalez-Prelcic, R. W. Heath, and A. Ghosh, "Leveraging sensing at the infrastructure for mmWave communication," *IEEE Commun. Mag.,* vol. 58, no. 7, pp. 84–89, 2020, doi: 10.1109/MCOM.001.1900700.

[13] S. Baek, D. Kim, M. Tesanovic, and A. Agiwal, "3GPP New Radio Release 16: Evolution of 5G for industrial Internet of Things," *IEEE Comm. Mag.,* vol. 59, no. 1, pp. 41–47, 2021, doi: 10.1109/MCOM.001.2000526.

[14] Y. Zhang, T. Arakawa, J. V. Krogmeier, C. R. Anderson, D. J. Love, and D. R. Buckmaster, "Large-scale cellular coverage analyses for UAV data relay via channel modeling," in *2020 IEEE Intl. Conf. Commun.*, Jun. 7–11, 2020, pp. 1–6, doi: 10.1109/ICC40277.2020.9149403.

[15] W. Khawaja, I. Guvenc, D. W. Matolak, U. Fiebig, and N. Schneckenburger, "A survey of air-to-ground propagation channel modeling for unmanned aerial vehicles," *IEEE Commun. Surveys Tuts.,* vol. 21, no. 3, pp. 2361–2391, 2019, doi: 10.1109/COMST.2019.2915069.

**YAGUANG ZHANG** [S'15] (ygzhang@purdue.edu) is a Post-Doctoral Research Assistant in Electrical and Computer Engineering at Purdue University.

**DAVID J. LOVE** [S'98, M'05, SM'09, F'15] (djlove@purdue.edu) is the Nick Trbovich Professor of Electrical and Computer Engineering at Purdue University.

**JAMES V. KROGMEIER** [M'89, SM'21] (jvk@purdue.edu) is a Professor of Electrical and Computer Engineering at Purdue University.

**CHRISTOPHER R. ANDERSON** [S'96, M'06, SM'11] (canderso@usna.edu) is an Associate Professor of Electrical Engineering at the United States Naval Academy.

**ROBERT W. HEATH** [S'96, M'01, SM'06, F'11] (rwheathjr@ncsu.edu) is the Distinguished Professor in the Department of Electrical and Computer Engineering at North Carolina State University.

**DENNIS R. BUCKMASTER** (dbuckmas@purdue.edu) is a Professor of Agricultural and Biological Engineering and the Dean's Fellow for Digital Agriculture at Purdue University.